\documentclass[preprint,10pt,a4paper]{aastex}
\usepackage[colorlinks,urlcolor=blue,citecolor=blue,linkcolor=blue]{hyperref}
\usepackage{amsmath}
\usepackage{graphicx}
\usepackage{multicol}
\newcommand{\be}{\begin{equation}} 
\newcommand{\ee}{\end{equation}}

\newcommand{\ba}{\begin{eqnarray}}
\newcommand{\ea}{\end{eqnarray}}

\newcommand\etal{\textit{et al.\ }}
\newcommand\eg{\textit{e.g.,\ }}

\newcommand{\Bf}{{magnetic field}}
\newcommand{\Bfs}{{magnetic fields}}

\newcommand{\NS}{neutron star}

\newcommand{\Ch}{Chandrasekhar}

\begin{document}

\title{Transients following white dwarfs mergers}

\author{Maxim Lyutikov\\
Department of Physics and Astronomy, Purdue University, 
 525 Northwestern Avenue,
West Lafayette, IN
47907-2036  \\
and\\
Silvia Toonen\\
Institute for Gravitational Wave Astronomy and School of Physics and Astronomy, University of Birmingham, Birmingham, B15 2TT, United Kingdom}

\begin{abstract}
Mergers of white dwarfs (WDs) may lead  to a variety of transient astrophysical events, SNIa being one possible outcome. 
Lyutikov \& Toonen (2017, 2019) argued that mergers of WDs result, under various parameter regimes,  in unusual central engine-powered supernova and a type of short Gamma Ray Bursts that show extended emission tails.  
Observations by Gvaramadze \etal\  (2019) of the central star and the nebula J005311 match  to the details the model of Lyutikov \& Toonen (2017, 2019)  for the  immediate product of a merger of a heavy ONeMg WD with CO WD (age, luminosity,  stellar size, hydrogen deficiency and chemical composition). 
 \end{abstract}

\section{Introduction: different channels of WD mergers}

Mergers of WDs is one of the most frequent catastrophic events, see \S \ref{rates} for discussion of the rates; \citep[see also][]{2012ApJ...748...35S,2016MNRAS.463.3461S}. Supernova of Ia type is the  channel  of WDs mergers most frequently discussed \citep[see ][for a recent review]{2014ARA&A..52..107M}. As we discuss  below, there are other various possible channels. \cite{2017arXiv170902221L,2019MNRAS.487.5618L} argued that under certain circumstances WDs mergers can lead to  other  transients -  from special types of SNe, like Fast Blue Optical Variables (FBOTs), to a specific type of Gamma Ray Bursts (short GRBs with extended tails)\footnote{ As discussed by \cite{Lyutikov:2009,2017arXiv170902221L},  there is  a number of observational  contradictions to the paradigm of short GRBs coming from the    binary \NS\ merger, like extended emission and late flares -- both not seen in the particular case of  GW/GRB170817).}.
 In particular, the model of 
\cite{2017arXiv170902221L,2019MNRAS.487.5618L} asserts  that the merger of a heavy ONeMg WD with a CO WD   creates a super-Chandrasekhar mass,  shell burning (C and O burning) star with  luminosity  $L\sim 10^4 L_\odot$.  The star  produces strong, nearly hydrogen deficient winds.

This picture was well matched by the  recent observations  of \cite{2019Natur.569..684G} who discovered  a hot, $\sim 200,000$K, luminous   $\sim10^{4.5} L_\odot$ star within a mid-infrared nebula. Both the star and the nebula appear to be free of hydrogen and helium. The wind velocity is exceptionally high, $\sim 16,000$ km s$^{-1}$.
\cite{2019Natur.569..684G} suggested that the central star and the nebula J005311 are  the system created by a WD merger;  \cite{2017arXiv170902221L,2019MNRAS.487.5618L} described a particular channel of WDs merger that may lead to J005311. In particular,  \cite{2019Natur.569..684G} discuss a possibility of a high neon surface abundance  that could  imply that a ONeMg WD participated in the merger event - also matching the model of   \cite{2017arXiv170902221L,2019MNRAS.487.5618L}.
 (Sakurai's Object \citep[\eg][]{1999A&A...343..507A} and R Coronae Borealis  \citep{ 1984ApJ...277..355W,2012ApJ...757...76S} may be other examples of peculiar shell-burning  stars following   white dwarf mergers .)

\section{Specific channel of collapse following  WDs mergers}

The fate of  WDs mergers mostly  depends on  masses of the components, which in turn depend on the initial zero-age Main-Sequence  masses and the initial orbital separation. They also depend on the less certain properties of the evolution of the contact binaries, like mass ejected {and orbital evolution} during the Common Envelope phases.

 The model of \cite{2017arXiv170902221L,2019MNRAS.487.5618L}  suggests that 
  an initial binary with components  of $\sim 5 M_\odot$ and  $\sim 8 M_\odot$ forms a close WD binary of a ONeMg and CO WDs, following two distinct evolutionary channels. The gravitational waves-induced merger results in  disruption of a lower massive WD, its accretion on the primary, avoiding  SN Ia-type explosion.
  Immediately after the merger the core (the initial primary) is surrounded by the shell and the disk made of the disrupted WD \citep{1990ApJ...348..647B}.
    The merger  produces an optical transient, possibly helped by the magnetic dissipation  \citep[e.g.][]{2014MNRAS.438..169B}.
  
   After the transients settles down, the  result is  a special  type shell-burning star with a  size few times $10^9$ cm,  fast rotating (at the surface), with luminosity $L\sim 10^4 L_\odot$, producing nearly hydrogen-clear  winds.     The star lives for $\sim  10^4$ years, while the  envelope mass is  both lost to the wind and added to the core as nuclear ashes. 
If the mass of the core exceeds the Chandrasekhar mass, the
electron-capture collapse follows.

The collapse produces various observed phenomena that depend on the particular properties of the merging system. In particular,
the observed properties of the collapse depend  on (i) duration of shell burning affecting the amount of envelope mass left at the moment of collapse; (ii) duration of shell burning  and the corresponding amount of  angular moment transferred to the core; (iii)   the viewing angle with respect to the axis. 


Duration of the shell burning  stage and the amounts of mass left in the  shell  at the moment of collapse, as well as  the spin of the core, strongly depends on how close the mass of ONeMg WD was to the \Ch\ mass before the merger. If the mass of the ONeMg WD  was close to the \Ch\ mass the collapse occurs with fairly large envelope mass, $\sim 0.5 m_\odot$, while the core was not spun up. 
For longer shell burning stage, envelopes are smaller, cores rotate faster. 
Mild envelope masses of few $10\% $ of $M_\odot$ result in {bright} short SNe, FBOTs. 
In special  cases when the mass of the envelope left at the moment of  collapse is very small,  SNe  like AT2018cow result. Finally, the shell may be fully expelled with the wind before the core reaches the \Ch\ mass. In this case a high mass WD is left behind \citep[hydrogen and helium deficient WD H1504+65 could be an example][]{2015A&A...584A..19W}

 The longer is the shell burning stage, the more angular momentum is transferred from the shell to the core. The details of the core collapse then depend on its total angular momentum:
high core angular momentum result in the formation of accretion disk around the newly formed NS, while smaller amount of angular momentum can allow direct collapse into a \NS,  followed by a bounce.

 \begin{figure}[h!]
\centering
\includegraphics[width=.99\textwidth]{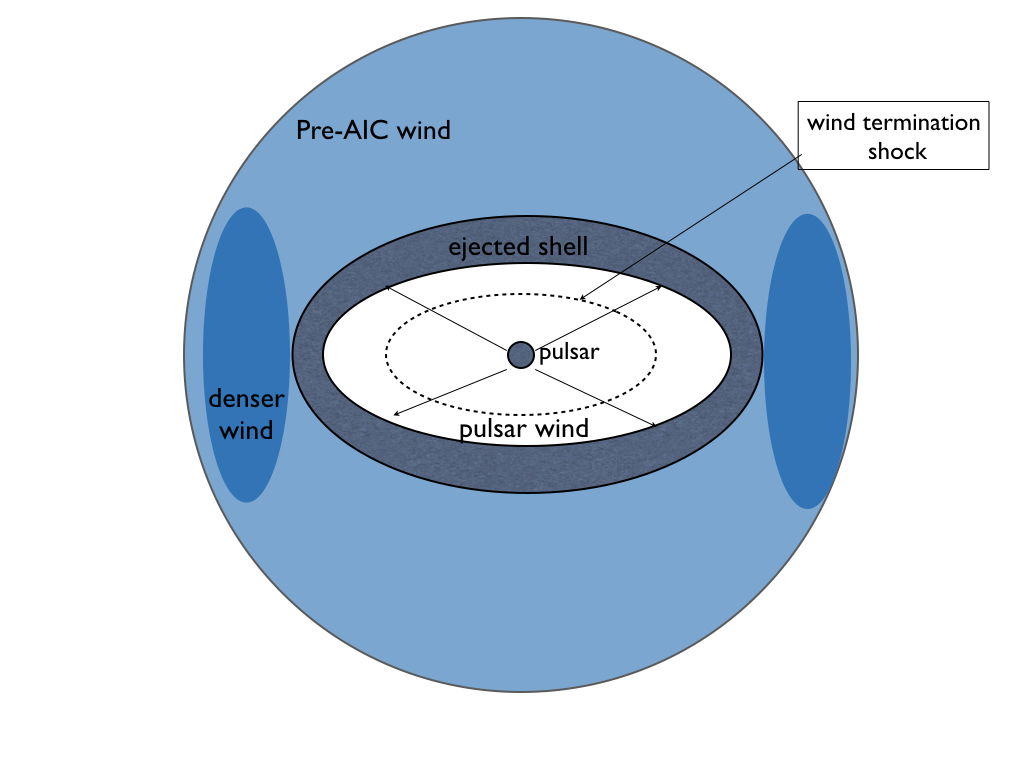}
\caption{Immediate surrounding of the system following the collapse \protect \citep[after] []{2019MNRAS.487.5618L}. The remaining NS generates  anisotropic pulsar-like wind (highly relativistic, highly magnetized with power $\sim \sin^2 \theta$), that interacts with the bounced-ejected shell, and the preceding  wind from the shell burning stage.  The pulsar winds,  the ejected shell and the pre-explosion wind are  all expected to be equatorially collimated.  }
\label{SurrondingSGRB} 
\end{figure}

\section{Magnetic fields and the central engine}

In most cases  of WDs mergers and the following collapse,  the resulting  central \NS\ plays the most  important role in 
powering of the observed   emission - as argued for the short GRB case by  \cite{2017arXiv170902221L} and for FBOTs  by  \cite{2019MNRAS.487.5618L} \citep[see also][]{2014ApJ...794...23D,2019ApJ...871...73H,2019ApJ...872...18M}. During the collapse,  the \Bf\ is amplified  \citep{2015Natur.528..376M},  and the \NS\ is spun to millisecond periods. As a result,  the newly  born  spinning \NS\   will produce  a long lasting highly magnetized relativistic wind
 that shocks against the ejecta (and,  later on,  against the pre-explosion wind). The NS wind-ejecta interaction will produce two shocks: forwards shock in the ejecta and termination shock in the wind.
 It is the wind termination shock that produces the X-ray  emission, while the radiation-dominated forward shock produces the optical transient (in the case of AT2018cow).
  In the termination shock the accelerated particles will produce synchrotron emission  in the fast cooling regime, so that a large fraction of the wind power will be emitted as radiation. \citep[See][for discussion of emission produced at the highly magnetized termination shock in GRBs]{2017ApJ...835..206L}.

 Powering by the central star through \Bfs\ is inferred in the case of J005311 by \cite{2019Natur.569..684G}. Here the \Bfs\ should be amplified in the burning shell \citep{
2013ApJ...773..136J}, while the post-collapse activity of the  central engine is powered by the \Bfs\ in the collapse core.

\section{The rates and the hosts}
\label{rates}

We estimated the rate of these events in \cite[] []{2019MNRAS.487.5618L} with a population synthesis approach using SeBa \citep{1996A&A...309..179P,2012A&A...546A..70T}. For a full description of the models, see the before-mentioned paper, here we provide a summary of the main results. 

We predict that for every solar mass of formed stars, there are  of the order of $(3-9)\cdot10^{-5}$ mergers between a ONe WD and a CO WD within a Hubble time. This is consistent with our estimate of the rates of FBOTs.  Here we adopt the estimate from \cite{2014ApJ...794...23D} that for every 100 core-collapse supernova there are 4-7 FBOTs. Furthermore, we take a core-collapse rate 0.0025-0.010M$_{\odot}^{-1}$ based on \citep{Mao17, Gra17}. 

For a galaxy like the Milky Way, with a constant star formation history of $4M_{\odot}$ yr$^{-1}$ for a Hubble time, the current merger rate of CO-ONe WDs is of the order of $(1-4)\times 10^{-4}$yr$^{-1}$. This is in agreement with the BPS calculations of \cite{Yun17} and \cite{2019MNRAS.484..698R}. CO-CO WD mergers have a merger rate of about $5\times 10^{-3}$yr$^{-1}$, and those with a combined mass  above the Chandrasekhar mass have a merger rate of $1.7-2.2\times 10^{-3}$yr$^{-1}$  in our simulations. 
Thus, only several percents of WD-WD mergers exceeding the Chandrasekhar mass produce an unusual transient, FBOTs or GRBs.

\cite{2019MNRAS.487.5618L} also estimated that the he distributions of delay times of the CO-ONeMg mergers after a single burst of starformation peaks at short delay times of about $\sim$ 50-100Myr, with a long tail to long delay times. Thus, there is a slight preference for  the host galaxies of CO-ONeMg mergers to be more similar to those of   core-collapse supernovae.

\section{Conclusion}

In this contribution we described how merger of WDs can produce peculiar SN-type events and special types of GRBs. We also mention here   that Type Ibn supernova
PS1-12sk that occurred in a galaxy  with low star formation rate
\citep{2013ApJ...769...39S,2019ApJ...871L...9H} could have followed a similar path, but involved a He WD.

We thank Vasilii Gvaramadze for comments on the manuscript.

\bibliographystyle{apj}

  \bibliography{/Users/maxim/Home/Research/BibTex}

\end{document}